\documentclass{moriond}
\newcommand{\Photo}{}

\def\be{\begin{equation}}
\def\ee{\end{equation}}
\def\bea{\begin{eqnarray}}
\def\eea{\end{eqnarray}}

\usepackage{tikz}
\usetikzlibrary{arrows.meta, positioning}
\usepackage{amsmath}
\usepackage{natbib}

\usepackage{xcolor}

\begin{document}
\vspace*{4cm}
\title{Study of cosmic ray impacts on cryogenic high sensitivity detectors}

\author{Anaïs Besnard, Valentin Sauvage, Bruno Maffei}

\address{Institut d'Astrophysique Spatiale (IAS), Université Paris-Saclay, Orsay, France}

\maketitle\abstracts{After the Planck mission’s launch in 2009, bolometers of its High Frequency Instrument (HFI) were considerably affected by cosmic rays, which necessitated several years of post-treatment to clean the data. To study the susceptibility of high sensitivity cryogenic detectors to particle impacts, IAS has developed the DRACuLA facility. We present the results of the latest two test campaigns performed on new generation of detectors.}

\section{Introduction}

After the Planck mission’s launch in 2009, the bolometers of its High Frequency Instrument (HFI) were significantly affected by cosmic rays, mainly composed of protons, alpha particles, and heavy nuclei with a mean energy $\approx$~10~GeV. Their interaction with the detectors or the focal plane induced energy depositions that produced intensity peaks in the data, referred to as ``glitches’’, thereby contaminating the astrophysical signal. More than 30\% of the data were affected by these events \cite{catalano_2014}, requiring several years of post-processing to remove them. As future balloon \cite{Masi_2010} and space missions aim for higher sensitivity through an increased number of detectors and larger focal planes, it is therefore essential to characterize the response of high-sensitivity cryogenic detector prototypes to particle hits prior to launch.

\section{Proton irradiation on cryogenic detectors}
With this aim, IAS has developped DRACuLA (Detector irRAdiation Cryogenic faciLity for Astrophysics), a customized cryogenic facility \cite{besnard_cryogenic_2024} to study the effects of particles impact on detector prototypes and/or part of focal planes, in their operating conditions, that can be coupled to particle accelerators. IAS has carried out three successful campaigns on three different type of detectors (bolometer, transition edge sensor and kinetic inductance detectors). We present here the two latest, with the main results and the glitches analysis performed for each.

\subsection{Transition edge sensors (TES)}
\begin{figure}[h!]
    \centering
    \begin{tikzpicture}[
      node distance = 0.75cm,
      box/.style = {
        rectangle,
        draw,
        rounded corners,
        minimum width  = 2.5cm,
        minimum height = 1.2cm,
        text centered,
        font = \small
      },
      arrow/.style = {
        -Stealth,
        thick
      }
    ]
    
    \node[box] (A) {DFT};
    \node[box, right=of A] (B) {Identify and filter stochastic noise};
    \node[box, right=of B] (C) {iDFT};
    \node[box, right=of C] (D) {offset};
    
    \draw[arrow] (A) -- (B);
    \draw[arrow] (B) -- (C);
    \draw[arrow] (C) -- (D);
    
    \end{tikzpicture}

    \caption{Procedure used to denoise a glitch through DFT: we use a DFT to then identify and filter the stochastic noise with a 10 kHz frequency cut. Reconstruction of the signal in the temporal domain with an inverse DFT and application of a temporal offset to fit to the two physical phenomena in the glitch's shape. This method allows to improve the fitting of the pulse shape \cite{tandem}. }
    \label{denoise}
\end{figure}
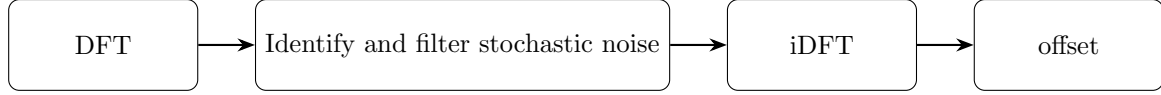

In 2024, we carried out an irradiation campaign to characterize TES prototypes designed by NIST for the LiteBIRD MHFT. We performed a single event characterization with protons at 18 and 22~MeV at ALTO – IJCLab in Orsay, France. From the $\approx 2800$ events we recorded during this campaign, we identified three different shapes of glitches, hypothetically triggered by protons hitting different locations of the detector chip. This very explorative work enabled us to extract the time constant with MCMC procedure, allowing us to fit the profile of the glitch and is therefore a crucial parameter to ease the cleaning of the data. We then denoised the data by filtering the stochastic noise through a discrete Fourier transform (DFT), allowing a better accuracy in the determination of the time constant, (see Fig.\ref{denoise}) \cite{tandem}.
\subsection{Kinetic inductance detectors (KIDs)}
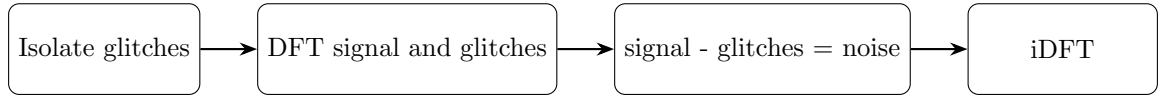
\begin{figure}[h!]
    \centering
    \begin{tikzpicture}[
      node distance = 0.75cm,
      box/.style = {
        rectangle,
        draw,
        rounded corners,
        minimum width  = 2.5cm,
        minimum height = 1.2cm,
        text centered,
        font = \small
      },
      arrow/.style = {
        -Stealth,
        thick
      }
    ]
    
    \node[box] (A) {Isolate glitches};
    \node[box, right=of A] (B) {DFT signal and glitches};
    \node[box, right=of B] (C) {signal - glitches = noise};
    \node[box, right=of C] (D) {iDFT};
    
    \draw[arrow] (A) -- (B);
    \draw[arrow] (B) -- (C);
    \draw[arrow] (C) -- (D);
    
    \end{tikzpicture}

    \caption{Deglitch of the signal : in the temporal domain, we isolate glitches and use a DFT on the signal and the glitches extracted. We then substract the glitches to the signal to obtain the deglitched signal. Finally, we reconstruct the deglitched signal with an inverse DFT.}
    \label{deglitch}
\end{figure}

In 2025, we carried out an irradiation campaign to characterize KID arrays designed and fabricated by SRON for PRIMA, NASA mission in phase A study. We performed a total dose proton irradiation (6 krad) at PARTREC – Groningen University, The Netherlands. This experiment allowed us to simulate 10 years of cosmic-rays while in operation at L2 point. In addition, DRACuLA’s inner components having been activated by the irradiation, around 130 glitches /s due to secondary particles were recorded during the post-irradiation measurements ($\approx$~15 \% of the dataset). Even after application of several deglitching methods, analysis is still in progress to deconvoluate low amplitude glitches from stochastic noise and to quantify the performances degradation, this will be described in \cite{prima}. \\
Until now we used the DFT only on indivdual glitches from the LiteBIRD campaign. We are applying the same method on a larger dataset such as the measurements done during the PRIMA campaign (see Fig.\ref{deglitch}). Finally, we will apply the procedure on subset of data of Planck HFI to validate it. 
\section{Conclusion}

In the context of future CMB balloon-borne and space missions, it is imperative to undertake a comprehensive study of particles’ impacts on new generation of highly sensitive detectors. The proton irradiation campaigns carried out on cryogenic high sensitivity detectors by IAS allows to bring answer to this problem. In addition, to enhance the reliability of the method to apply to the current data and ensure optimal performance in subsequent campaigns, a procedure using Fourier domain is currently being investigated, method which will be validated through application on Planck HFI data.

\bibliography{Bibliography}

\end{document}